\documentstyle[multicol,prl,aps]{revtex}

\begin{document}
\draft
\title{Superconducting 2D system with lifted spin degeneracy:\\
Mixed singlet-triplet state}
\author{Lev P. Gor'kov\cite{Gorkov*}}
\address{National High Magnetic Field Laboratory, Florida State University, Tallahassee, Florida 32310}
\author{Emmanuel I. Rashba\cite{Rashba*}}   
\address{Department of Physics, MIT, Cambridge, Massachusetts 02139}  
\date{\today}
\maketitle
\begin{abstract} 
Motivated by recent experimental findings, we have developed a theory of the superconducting state for 2D metals without inversion symmetry modeling the geometry of a surface superconducting layer in a field-effect-transistor or near the boundary doped by adsorbed ions. In such systems the two-fold spin degeneracy is lifted by spin-orbit interaction, and singlet and triplet pairings are mixed in the wave function of the Cooper pairs. As a result, spin magnetic susceptibility becomes anisotropic and Knight shift retains finite and rather high value at $T=0$.                
\end{abstract}
\begin{multicols}{2}
\narrowtext
In this presentation we consider the energy spectrum and some other properties of the superconducting state for a two-dimensional (2D) system without inversion symmetry when the two-fold degeneracy of the electron energy spectrum is lifted by spin-orbit (SO) interaction. We have been motivated by several recent experimental discoveries. Superconductivity (SC) with critical temperature $T_c = 91$ K was observed in insulating WO$_3$ when its surface was doped by Na$^+$ ions \cite{Reich,Shengelaya}. According to Refs.~\onlinecite{Reich,Shengelaya}, SC in the material is a surface effect which comes to the existence in a narrow layer with a composition of ${\rm Na}_{0.05}{\rm WO}_3$. We speculate that strong near-surface electric fields should be present in the SC layer of these crystals quite similar to the 2D superconducting fullerene and polyacene crystals in the field-effect transistor geometry \cite{Schon1,Schon2}. The two-fold degeneracy of the electron energy spectrum in the bulk characteristic of crystals with 3D inversion symmetry should be broken in the surface layer because of SO coupling. This effect is known and well documented experimentally \cite{Schultz,Nitta} in physics of semiconductor microstructures, and has been recently discovered in surface states of metallic Au \cite{LaShell} and also of (Li covered) Mo and W \cite{Rotenberg}. In what follows, therefore, we discuss general properties of 2D superconductors with lifted spin degeneracy. 

Ideal boundary being a strong ``defect'' for the crystal's bulk, it is known that split-off 2D electron bands may form at the surface (Tamm levels). Besides, near the surface the space inversion (the 3D inversion which ensures two-fold degeneracy of the bands throughout the Brillouin zone\cite{Elliott}) is automatically broken, resulting in a new SO invariant of the group C$_{\infty {\rm v}}$ (and its discrete sub-groups) in the free electron Hamiltonian
\begin{equation}
H_{\rm so} = \alpha ({\bbox \sigma}\times {\bf p})\cdot{\bf n}
\label{eq1}
\end{equation}
dubbed in the literature as Rashba term\cite{R60,BR84}. Here $\bf p$ is a 2D quasimomentum, 
$\bbox \sigma$ are Pauli matrices, and $\bf n$ is a unit vector normal to the surface 
($\hbar =1$). Despite the fact that $\bf p$ is two-dimensional, the 3D group controls the symmetry of the system since $\bf n$ determines the prefered direction of the  normal to the plane and spins are always 3D. Total single electron Hamiltonian, $H=H_0+H_{\rm so}$, diagonalizes in eigenfunctions $\eta_\lambda({\bf p})$ and eigenvalues $\varepsilon_\lambda(p)$
\begin{eqnarray}
\eta_\lambda ({\bf p})&=& {1\over \sqrt{2}}\left(
\begin{array}{c}
1\\i\lambda \exp(i\varphi_{\bf p})
\end{array}\right), \nonumber \\ 
\varepsilon_\lambda(p)=\varepsilon_0(p)&+&\varepsilon^{\rm so}_\lambda(p),\;\;
\varepsilon^{\rm so}_\lambda(p)=-\alpha\lambda p \;,
\label{eq2}
\end{eqnarray}
$\lambda =\pm 1$, and the Fermi surface splits into two sheets. SO interaction $H_{\rm so}$ puts electron spins into the plane, ($\perp{\bf n}$), where they are aligned perpendicular to the quasimomentum $\bf p$.  The magnitude of the coefficient $\alpha$ depends on the electric field, presence of adatoms at the boundary, atomic weight, and atomic shells involved\cite{PH00}. Angle resolved photoemission spectroscopy (ARPES) data on the {\it s-p} surface band of Au are in excellent agreement with Eq.~(\ref{eq1}). They result in a nearly isotropic spectrum with $\alpha \approx 5\times 10^{-9} ~{\rm eV}\cdot {\rm cm}$ (that is somewhat larger than in typical semiconductor systems) and in the SO splitting at the Fermi level $2\vert\alpha\vert p_F = 0.11$ eV\cite{LaShell}. Anisotropic $d$-derived bands of Mo and W show even larger SO splittings up to 0.13 eV and 0.5 eV, respectively, depending on the surface concentration of adsorbed Li$^+$ ions. 

Returning back to the surface SC reported in Refs.~\onlinecite{Reich,Shengelaya} for doped tungsten bronzes, we expect, therefore, that the SO interaction of Eq.~(\ref{eq1}) should play essential role in the surface properties both of the normal and SC phases. In what follows we address changes in properties of the BCS-type 2D superconductivity caused by lifting the electron spin degeneracy in the presence of the SO interaction.

For the simplicity sake we consider a two-particle interaction 
$U(\vert{\bf r}-{\bf r}^{\prime}\vert)$ (here and below all vectors lie in the plane). Retardation effects such as due to the phonon mediated attraction are known to be of no importance for a weak coupling scheme. To put it briefly, we do not address at all the issue of the mechanism responsible for pairing, and restrict ourselves by features pertaining to the reduced dimensionality (for the proper symmetry at the surface) and spin-split isotropic electron spectrum. The SO splitting at the Fermi level, $2\alpha p_F$, (here and below $\alpha >0$) can be comparable to, or exceed, the SC characteristic scale, the critical temperature $T_c$. However, it is assumed to be small compared to the Fermi energy, $2\alpha p_F\ll \varepsilon_F$, and 
$p_+ - p_-\ll p_F$, where $p_\lambda =p_\pm$ are Fermi momenta for the two SO split sheets of the Fermi surface. Therefore, near the Fermi energy the electron dispersion is taken linear:
\begin{equation}
\xi_\lambda(p) = v_F (p-p_\lambda), \;\; p_{\lambda}=(1+\alpha\lambda/v_F)p_F.
\label{eq3}
\end{equation}

The total electron Hamiltonian $H = H_0 + H_{\rm so} + H_{\rm int}$
includes the interaction Hamiltonian
\begin{eqnarray}
H_{\rm int} &=&\case 1/2\sum_{\lambda\mu\nu\rho}\sum_{{\bf p}{\bf p}^\prime{\bf q}}
U_{\lambda\mu\nu\rho}({\bf p},{\bf p}^\prime,{\bf q})\nonumber \\
&\times&a_\lambda^+({\bf p})a_{\mu}^+(-{\bf p}-{\bf q})
a_{\nu}(-{\bf p}^\prime-{\bf q})a_\rho({\bf p}^\prime).
\label{eq5}
\end{eqnarray}
Here $a_\lambda({\bf p})$ are Fermi electronic operators, and 
$U_{\lambda\mu\nu\rho}({\bf p},{\bf p}^\prime,{\bf q})$ is the interaction potential written in the representation of the spinors $\eta_\lambda({\bf p})$
\begin{eqnarray}
U_{\lambda\mu\nu\rho}({\bf p}&,&{\bf p}^\prime,{\bf q})=
U(\vert{\bf p}-{\bf p}^\prime\vert)\nonumber \\
&\times&(\eta_\lambda({\bf p}), \eta_\rho({\bf p}^\prime))
(\eta_\mu(-{\bf p}-{\bf q}), \eta_\nu(-{\bf p}^\prime-{\bf q})),
\label{eq6}
\end{eqnarray}
$U(\vert{\bf p}\vert)$ is the Fourier image of the interaction potential, and scalar products of spinors are equal to
\begin{equation}
(\eta_\lambda({\bf p}), \eta_{\lambda^\prime}({\bf p}^\prime))
= \case 1/2 \{1+\lambda \lambda^\prime\exp[-i(\varphi_{\bf p}-\varphi_{{\bf p}\prime})]\}.
\label{eq7}
\end{equation}

In what follows we employ the method of thermodynamic (Matsubara) Green functions\cite{AGD}.
Onset of the SC state manifests itself as the appearance, side by side with the regular Green functions\cite{nondiag}
\begin{equation}
g_\lambda({\bf p}, \tau-\tau^\prime)=-\langle T\{{\tilde a}_\lambda({\bf p}, \tau)
{\tilde a}^+_\lambda({\bf p}, \tau^\prime)\}\rangle,
\label{eq8}
\end{equation}
also of the non-zero anomalous (Gor'kov) averages\cite{Gorkov}
\begin{equation}
f_\lambda({\bf p}, \tau-\tau^\prime)=\lambda\langle T\{{\tilde a}_\lambda({\bf p}, \tau)
{\tilde a}_\lambda(-{\bf p}, \tau^\prime)\}\rangle,
\label{eq9}
\end{equation}
\begin{equation}
f_\lambda^+({\bf p}, \tau-\tau^\prime)=\lambda\langle T\{{\tilde a}^+_\lambda({\bf p}, \tau)
{\tilde a}^+_\lambda(-{\bf p}, \tau^\prime)\}\rangle,
\label{eq10}
\end{equation}
where operators ${\tilde a}_\lambda({\bf p}, \tau)$ are in the Heisenberg representation. 
Functions $f_\lambda({\bf p})\equiv f_\lambda({\bf p}, 0+)$ that obey the relation
\begin{equation}
f^{*}_\lambda({\bf p})=f^+_\lambda(-{\bf p})
\label{eq11}
\end{equation}
are components of the wave functions of Cooper pairs at each spectrum branch. Fermionic commutation relations result in
\begin{equation}
f_\lambda({\bf p})=-f_\lambda(-{\bf p}).
\label{eq12}
\end{equation}
From equations of motion for the operators ${\tilde a}_\lambda({\bf p},\tau)$, 
$\partial_\tau {\tilde a}_\lambda({\bf p}, \tau)=[H, {\tilde a}_\lambda({\bf p},\tau)]$, equations of motion for $g_\lambda({\bf p}, \tau-\tau^\prime)$ and $f_\lambda({\bf p}, \tau-\tau^\prime)$ follow in the usual way. They include 4-fermion $T$-products, the standard decoupling\cite{AGD,Gorkov} of which results in
\end{multicols} 
\widetext
\begin{equation}
(\partial_\tau +\varepsilon_\lambda(p)) g_\lambda({\bf p}, \tau -\tau^\prime)-
\lambda\sum_{\mu {\bf p}^\prime}\mu U_{\lambda\lambda\mu\mu}({\bf p},{\bf p}^\prime, 0)
f_\mu(-{\bf p}^\prime)f_\lambda^+(-{\bf p}, \tau -\tau^\prime)
=-\delta(\tau -\tau^\prime),
\label{eq13}
\end{equation}
\begin{equation} 
(\partial_\tau -\varepsilon_\lambda(p)) f_\lambda^+(-{\bf p}, \tau -\tau^\prime)+
\lambda\sum_{\mu {\bf p}^\prime}\mu U_{\mu\mu\lambda\lambda}({\bf p}^\prime,{\bf p}, 0)
f^+_\mu({\bf p}^\prime)g_\lambda({\bf p}, \tau -\tau^\prime)=0.
\label{eq14}
\end{equation}
\begin{multicols}{2}
\narrowtext
For the Cooper pairs with the zero total momentum forming the condensate, ${\bf q}=0$ in 
$U_{\lambda\lambda\mu\mu}({\bf p},{\bf p}^\prime,{\bf q})$. Next simplification in the potential comes from the fact that it enters in Eqs.~(\ref{eq13},\ref{eq14}) only in the combination with the functions $f_\mu({\bf p})$, hence, because of Eq.~(\ref{eq12}), 
$U_{\lambda\lambda\mu\mu}({\bf p},{\bf p}^\prime,0)$ should be antisymmetrisized. Simple algebra shows
\begin{eqnarray}
U_{\lambda\mu}({\bf p},{\bf p}^\prime)&\equiv &
 \case 1/2[U_{\lambda\lambda\mu\mu}({\bf p},{\bf p}^\prime,0)-
U_{\lambda\lambda\mu\mu}({\bf p},-{\bf p}^\prime,0)] \nonumber \\
&\approx &\case 1/2 \lambda\mu U(0) \exp[i(\varphi_{{\bf p}^\prime}-\varphi_{\bf p})].
\label{eq15}
\end{eqnarray}
Having in mind $s$-pairing we neglected the momentum dependence of the potential in the right hand side of Eq.~(\ref{eq15}). Making Fourier transformation of Eqs.~(\ref{eq13},\ref{eq14})  
to the Matsubara frequencies $i\omega_n$, we find the final form of Gor'kov equations
\begin{equation}
[i\omega_n -\varepsilon_\lambda({\bf p})]g_\lambda({\bf p},\omega_n)+
\Delta({\bf p})f^+_\lambda(-{\bf p},\omega_n)=1,
\label{eq16}
\end{equation}
\begin{equation}
\Delta^+({\bf p})g_\lambda({\bf p}, \omega_n)+
[i\omega_n+\varepsilon_\lambda({\bf p})]f^+_\lambda(-{\bf p},\omega_n)=0,
\label{eq17}
\end{equation}
where the ``gap'' $\Delta({\bf p})$ is  
\begin{eqnarray}
\Delta({\bf p})=\exp(-i\varphi_{\bf p}){\Delta_0},&& 
\Delta_0=\case 1/2U(0)\sum_{\lambda {\bf p}}\exp(i\varphi_{\bf p})f_\lambda({\bf p}),
\nonumber \\
 \Delta^+({\bf p})&=&\Delta^{*}({\bf p}).
\label{eq18}
\end{eqnarray}
Before turn to the discussion of the nature of the order parameter in the new SC state, let us make one preliminary comment. The ``gap function'' $\Delta({\bf p})$ depends on $\bf p$ through its phase. This dependence is inherent in the non-perturbative character of the spinor basis functions of Eq.~(\ref{eq2}) after spin degeneracy is lifted. It cannot be fully eliminated, however, it can be changed by a different choice of phase factors in Eq.~(\ref{eq2}). Green functions diagonal in branch indices \cite{nondiag} are
\begin{eqnarray}
g_\lambda({\bf p}, \omega_n)&=&-[i\omega_n +\xi_\lambda({\bf p})]/D_{n\lambda}({\bf p}),
\nonumber \\
 f^+_\lambda(-{\bf p},\omega_n)&=&\Delta^+({\bf p})/D_{n\lambda}({\bf p}), \nonumber \\
D_{n\lambda}({\bf p})&=& \omega_n^2+\xi_\lambda^2({\bf p})+\vert\Delta\vert^2.
\label{eq19}
\end{eqnarray}
The self-consistency condition for $\vert\Delta ({\bf p})\vert\equiv\Delta(T)$ is
\begin{equation}
-\case 1/2 T U(0) \sum_{n\lambda{\bf p}}D^{-1}_{n\lambda}({\bf p})=1.
\label{eq20}
\end{equation}

In the approximation $2\alpha p_F\ll \varepsilon_F$, left hand side of Eq.~(\ref{eq20}) does not depend on $\alpha$ and coincides with the BCS equation for $\alpha =0$. The energy spectrum found from the poles of $g_\lambda({\bf p},\omega_n)$ and $f_\lambda({\bf p},\omega_n)$ with the proper account of Eq.~(\ref{eq3}) is
\begin{equation}
E_\lambda(p) = \pm [v_F^2(p-p_\lambda)^2+\Delta^2(T)]^{1/2},
\label{eq21}
\end{equation}
i.e., it consists of two gapped branches. One sees that weak SO coupling makes no changes in thermodynamical properties of the new BCS like state. In particular, the relation 
$\Delta (0) =(\pi/\gamma)T_c\approx 1.76 T_c$ holds in this case.

Green function were found above in the representation of energy branches indices $\lambda$. To get insight on the symmetry of the new SC state we rewrite them as matrices 
${\hat G}({\bf p},\omega_n)$ and ${\hat F}({\bf p},\omega_n)$ in the basis of the components 
$\eta^\alpha_\lambda({\bf p})$ of the spinors $\eta_\lambda({\bf p})$
\begin{eqnarray}
G_{\alpha\beta}({\bf p},\omega_n)&=&
\sum_\lambda\eta_\lambda^\alpha({\bf p})g_\lambda({\bf p},\omega_n)
\eta_\lambda^\beta({\bf p})^*,\nonumber \\
F_{\alpha\beta}({\bf p},\omega_n)&=&\sum_\lambda\eta_\lambda^\alpha({\bf p})
f_\lambda({\bf p},\omega_n)\eta_\lambda^\beta({\bf p}),
\label{eq22}
\end{eqnarray}
and split $g_\lambda({\bf p},\omega_n)$ and $f_\lambda({\bf p},\omega_n)$ into their symmetric and antisymmetric in $\lambda$ parts
\begin{equation}
g_\lambda = g_{\rm s} +\case 1/2 \lambda g_{\rm as}, \;\;
f_\lambda = f_{\rm s} +\case 1/2 \lambda f_{\rm as}.
\label{eq23}
\end{equation}
Eq.~(\ref{eq23}) implies that ${\hat G}={\hat G}_{\rm s}+{\hat G}_{\rm as}$ and 
${\hat F}={\hat F}_{\rm s}+{\hat F}_{\rm as}$. Antisymmetric parts of $g_\lambda$ and 
$f_\lambda$ (and also of ${\hat G}$ and ${\hat F}$) originate because of the splitting of the energy spectrum by the SO interaction. It is convenient to write ${\hat G}_{\rm s}$ and 
${\hat G}_{\rm as}$ (and also ${\hat F}_{\rm s}$ and ${\hat F}_{\rm as}$) in terms of the Pauli matrices $\bbox \sigma$ and the unit matrix $\sigma_0$. After some algebra we get
\begin{equation}
{\hat G}_{\rm s}({\bf p},\omega_n)=g_{\rm s}({\bf p},\omega_n)\sigma_0,
\label{eq24}
\end{equation}
\begin{equation}
{\hat G}_{\rm as}({\bf p},\omega_n)
= \case 1/2 g_{\rm as}({\bf p},\omega_n)({\bf p}^0 \times{\bbox \sigma})\cdot{\bf n},
\label{eq25}
\end{equation}
\begin{equation}
{\hat F}_{\rm s}({\bf p},\omega_n)=
[ f_{\rm s}({\bf p},\omega_n){\rm e}^{i\varphi_{\bf p}} ] \sigma_y,
\label{eq26}
\end{equation}
\begin{equation}
{\hat F}_{\rm as}({\bf p},\omega_n)=
-[f_{\rm as}({\bf p},\omega_n){\rm e}^{i\varphi_{\bf p}}]\sigma_y ({\bbox \sigma}\cdot{\bf p}^0),
\label{eq27}
\end{equation}
where ${\bf p}^0={\bf p}/p$. One sees that phase factors ${\rm e}^{i\varphi({\bf p})}$ drop out from both ${\hat F}_{\rm s}$ and ${\hat F}_{\rm as}$ according to Eqs.~ (\ref{eq18},\ref{eq19}). ${\hat G}_{\rm as}$ and ${\hat F}_{\rm as}$ components of the Green functions are of the order of unity but only inside a narrow ring of the momentum space, $\Delta p\approx p_+-p_-$, between two Fermi surfaces.

Both contributions to $\hat G$ are invariants of the group C$_{\infty {\rm v}}$, however, they are of opposite parity if the transformation ${\bf p}\rightarrow -{\bf p}$, rotation by $\pi$ around the $z$ axis, is performed only in coordinate space with spin remaining unchanged. Both 
${\hat G}_{\rm as}$ and ${\hat F}_{\rm as}$ depend on the azimuth of ${\bf p}^0$. The factor $({\bbox \sigma}\cdot{\bf p}^0)$ is a pseudo-scalar of the 3D inversion group. We attribute to ${\hat G}_{\rm as}$ a number of transport effects in normal conductors with lifted spin degeneracy which have been observed or predicted.

Existence of non-zero $\hat F$-function, Eqs.~(\ref{eq26}) and (\ref{eq27}), means the broken gauge symmetry in the SC state, $U(1)$. Eq.~(\ref{eq12}), when expressed in terms of the functions $F_{\alpha\beta}({\bf p}, \omega_n)$, reflects permutational symmetry of the wave function of the two electrons of the pair, $({\bf p},\alpha)$ and $(-{\bf p},\beta)$. If there is the 3D inversion center, one can make the total wave function odd either by choosing an antisymmetric spin wave function while the space part is symmetric (singlet pairing, $S=0$), or, {\it vice versa}, by choosing a symmetric spin wave function while the space part is antisymmetric (triplet pairing) ($S=1$), see Ref.~\onlinecite{inversion}. Therefore, 
$F_{\alpha\beta}\propto (\sigma_y)_{\alpha\beta}$ for $S=0$ and 
$F_{\alpha\beta}\propto (\sigma_y({\bbox\sigma}\cdot {\bf p}))_{\alpha\beta}$ for $S=1$. For the broken 3D inversion near the surface, {\it the Cooper pair wave function $\hat F$ becomes a mixture of singlet and triplet pairings}. ${\hat F}_{\rm s}({\bf p},\omega_n)$ is the singlet part component, while ${\hat F}_{\rm as}({\bf p},\omega_n)$ provides for the triplet admixture. Note that this term comes about even for the short range interaction 
$U(\vert{\bf r}-{\bf r}^\prime\vert) \propto \delta ({\bf r}-{\bf r}^\prime)$.

As an example of the phenomenon in which breaking the parity of the order parameter manifests itself, we calculate the spin susceptibility $\chi^{\rm sp}(T)$ of the surface SC state. It can be directly tested by Knight shift experiments. For systems with strong SO coupling, the total magnetic susceptibility $\chi$ cannot be split into the orbital and spin parts. However, when $\alpha\ll v_F$ such a division is justified\cite{Boiko}, and we find $\chi^{\rm sp}$ neglecting the quantization of the orbital motion. Following Ref.~\onlinecite{AbGor}, we calculate the tensor $\chi^{\rm sp}_{ij}$ ($i,j=x,y,z$) as a linear spin response to the magnetic field 
$\bf B$ 
\end{multicols}
\widetext 
\begin{equation}
\chi^{\rm sp}_{ij}= -\mu^2_{\rm B} T\nu(\varepsilon_F)\sum_{\omega_n}\int d\xi\ 
{\rm tr}\{\sigma_i {\hat G}({\bf p},\omega_n) \sigma_j {\hat G}({\bf p},\omega_n) -
\sigma_i {\hat F}({\bf p},\omega_n) \sigma_j {\hat F}^+({\bf p},\omega_n)\}.
\label{eq28}
\end{equation}
\begin{multicols}{2}
\narrowtext
\noindent
Here the trace is taken over the spin indices, $\nu(\varepsilon_F)=p_F/2\pi v_F$ is the density of states at the Fermi level (per one spin orientation), and $\mu_{\rm B}$ is the Bohr magneton. In the normal state the right hand side of Eq.~(\ref{eq28}) diverges because of Landau singularity\cite{AGD}, and to resolve it the summation over $\omega_n$ should be performed first. Finally
\begin{equation}
\chi_N^{\rm sp} = 2\mu^2_{\rm B} \nu(\varepsilon_F).
\label{eq29}
\end{equation}
Remarkably, $\chi_N^{\rm sp}$ found in the weak field limit, $\mu_{\rm B}B\ll \alpha p_F$, when Zeeman splitting is small compared to the SO splitting, is isotropic and coincides exactly with the Pauli susceptibility. Since the difference in $\chi^{\rm sp}$ in the normal and SC states comes from the integration over the region of only about $\Delta(T)$ near $\varepsilon_F$, the appropriate integral converges. Therefore, in these terms the integration  over $\xi$ can be performed first. Non-diagonal components of $\chi^{\rm sp}$ are equal zero because of the C$_{\infty {\rm v}}$ symmetry, while for the diagonal components of $\chi^{\rm sp}$ somewhat cumbersome calculations result in 

\end{multicols}
\widetext 
\begin{equation}
\chi^{\rm sp}_\perp(T) = \chi_N^{\rm sp}\left\{1 - \pi T\sum_{\omega_n}
{{\Delta^2(T)}\over {\sqrt{\omega_n^2+ {\Delta^2(T)}}}}
\cdot {1\over \omega_n^2+ \Delta^2(T)+(\alpha p_F)^2}  \right\}
\label{eq30}
\end{equation}

\begin{equation}
\chi^{\rm sp}_\parallel(T) = \case 1/2 \chi^{\rm sp}_\perp(T) +
\case 1/2  \chi_N^{\rm sp}\left\{1 - \sum_{\omega_n}
{{\pi T\Delta^2(T)}\over {\left(\omega_n^2+ {\Delta^2(T)}\right)^{3/2}}}
\right\}.
\label{eq31}
\end{equation}

\begin{multicols}{2}
\narrowtext
\noindent
It follows from Eqs.~(\ref{eq30}) and (\ref{eq31}) that 
$\chi^{\rm sp}_\perp(T)\neq \chi^{\rm sp}_\parallel(T)$ and $\chi^{\rm sp}_\perp(0)$, 
$\chi^{\rm sp}_\parallel(0) \neq 0$, i.e., {\it spin succeptibility is anisotropic and does not turn into zero at $T=0$}. We attribute this fact to the admixture of the triplet state to the $s$-type SC ground state because of the broken parity. For $\alpha p_F\ll\Delta(T)$, it follows from Eqs.~(\ref{eq30},\ref{eq31}) that 
$\chi^{\rm sp}_\perp(0)\approx{\case 2/3}[\alpha p_F/\Delta(T)]^2\chi_N^{\rm sp}$ and 
$\chi^{\rm sp}_\parallel(0)\approx {\case 1/3}[\alpha p_F/\Delta(T)]^2\chi_N^{\rm sp}$. Hence, for $\alpha =0$ the BCS result $\chi^{\rm sp}_\perp(0)=\chi^{\rm sp}_\parallel(0)=0$ is recovered. In the opposite 
limit $\Delta(T)\ll\alpha p_F$, we get $\chi^{\rm sp}_\perp(0)\approx \chi_N^{\rm sp}$ and $\chi^{\rm sp}_\parallel(0)\approx \chi_N^{\rm sp}/2$, hence, $\chi^{\rm sp}_\perp(T)$ is nearly $T$ independent while $\chi^{\rm sp}_\parallel(T)$ drops twice when $T$ changes from $T_c$ to $T=0$.

In conclusion, we have shown that spin-orbit interaction lifts the spin degeneracy for 2D (surface) superconductor resulting in the two gapped branches in the energy spectrum. Thermodynamics of such a state would be almost identical to a BCS like superconductor. It turns out that due to broken 3D inversion symmetry at the surface, the pair wave function is the mixture of singlet and triplet components. To the best of our knowledge, such a mixture has never been studied before. As the result, strong anisotropy appears in the spin-susceptibility tensor and the Knight shift. Splitting of two gapped branches in the momentum space depends on the strength of SO coupling. The latter can considerably exceed the SC scale $T_c$ to be observed by other means.

L. P. G. thanks Z. Fisk for discussions of a possibility to observe surface SC in tungsten bronzes in the field-effect-transistor geometry. His work was supported by the National High Magnetic Field Laboratory through the NSF Cooperative agreement No. DMR-9521035 and the state of Florida.

\end{multicols} 

\begin{references}
\bibitem[*]{Gorkov*} Also at L. D. Landau  Institute for Theoretical Physics, Russian Academy of Sciences, 117334 Moscow.
\bibitem[\dagger]{Rashba*} erashba@mailaps.org.
\bibitem{Reich} S. Reich and Y. Tsabba, Eur. Phys. J. B {\bf 9}, 1 (1999).
\bibitem{Shengelaya} A. Shengelaya, S. Reich, Y. Tsabba, and A. M\"{u}ller, 
Eur. Phys. J. B {\bf 12}, 13 (1999).
\bibitem{Schon1} J. H. Sch\"{o}n, Ch. Kloc, R. C. Haddon, and B. Battlog, 
Science {\bf 288}, 656 (2000).
\bibitem{Schon2} J. H. Sch\"{o}n, Ch. Kloc, and B. Battlog,
Nature {\bf 406}, 702 (2000).
\bibitem{Schultz} M. Schultz, F. Heinrichs, U. Merkt, T. Colin, T. Scauli, and S. L{\o}vold,
Semicond. Sci. Technol. {\bf 11}, 1168 (1996).
\bibitem{Nitta} J. Nitta, T. Akazaki, H. Takayanagi, and T. Enoki,
Phys. Rev. Lett. {\bf 78}, 1335 (1997).
\bibitem{LaShell} S. LaShell, B. A. McDougall, and E. Jensen, Phys. Rev. Lett. 
{\bf 77}, 3419 (1996).
\bibitem{Rotenberg} E. Rotenberg, J. W. Chung, and S. D. Kevan, Phys. Rev. Lett.
{\bf 82}, 4066 (1999).
\bibitem{Elliott} R. J. Elliott, Phys. Rev. {\bf 96}, 280 (1954).
\bibitem{R60} E. I. Rashba, Sov. Phys. - Solid State, {\bf 2}, 1109 (1960).
\bibitem{BR84} Yu. A. Bychkov and E. I. Rashba, Sov. Phys. - JETP Lett.
{\bf 39}, 78 (1984). For review see E. I. Rashba and V. I. Sheka, in: {\it Landau Level Spectroscopy}, ed. by G. Landwehr and E. I. Rashba (North-Holland, Amsterdam, 1991), 
v. 1, p. 131.
\bibitem{PH00} L. Petersen and P. Hedeg{\aa}rd, Surf. Sci. {\bf 459}, 49 (2000).
\bibitem{AGD} A. A. Abrikosov, L. P. Gor'kov, and I. E. Dzyaloshnskii, {\it Methods of Quantum
Field Theory in Statistical Physics}, Dover, N.Y. (1975).
\bibitem{nondiag} Green functions are diagonal in $\lambda$ for our short-range isotropic 
model, and may stay diagonal even under more general conditions. 
\bibitem{Gorkov} L. P. Gor'kov, Sov. Phys. - JETP {\bf 7}, 505 (1958).
\bibitem{inversion} G. E. Volovik and L. P. Gor'kov, JETP Lett. {\bf 39}, 674 (1984); 
Sov. Phys. - JETP {\bf 61}, 843 (1985). For rewiev see: L. P. Gor'kov, Sov. Sci. Rev. A Phys.
{\bf 9}, 1 (1987); M. Sigrist and K. Ueda, Rev. Mod. Phys. {\bf 63}, 239 (1991).
\bibitem{Boiko} I. I. Boiko and E. I. Rashba, Sov. Phys. - Solid State {\bf 2}, 1692 (1960).
\bibitem{AbGor} A. A. Abrikosov and L. P. Gor'kov, Zh. Eksp. Teor. Fiz. {\bf 42}, 1088 (1962).
\end{references}
\end{document}